\documentclass[runningheads]{llncs}

\usepackage{amsmath} 
\usepackage{amssymb}
\usepackage{bm} 
\usepackage[dvipdfmx]{graphicx, color} 
\usepackage{svg}
\usepackage{booktabs} 
\usepackage{tabularx} 
\usepackage{cite}
\usepackage{url}

\newcommand{\bs}[1]{\boldsymbol{#1}}

\newcommand{\argmax}{\mathop{\rm arg~max}\limits}

\begin{document}
\title{Bayesian Sparse Covariance Structure Analysis for Correlated Count Data}

\author{
Sho Ichigozaki\inst{1,2}\and
Takahiro Kawashima\inst{1} \and
Hayaru Shouno\inst{1}
}
%
\institute{
The University of Electro-Communications, Chofu, Tokyo, Japan\and
National Police Agency Info-Communications Bureau, Chiyoda-ku, Tokyo, Japan\\
\email{\{ichigozaki.show,shouno\}@uec.ac.jp}
}
\maketitle              
\begin{abstract}
In this paper, we propose a Bayesian Graphical LASSO for correlated countable data and apply it to spatial crime data. 
In the proposed model, we assume a Gaussian Graphical Model for the latent variables which dominate the potential risks of crimes.
To evaluate the proposed model, we determine optimal hyperparameters which represent samples better.
We apply the proposed model for estimation of the sparse inverse covariance of the latent variable and evaluate the partial correlation coefficients. 
Finally, we illustrate the results on crime spots data and consider the estimated latent variables and the partial correlation coefficients of the sparse inverse covariance.

\end{abstract}

\section{Introduction}
Revealing correlations of data is the simplest way to analyze relationships of given samples.
However, the simple way has a lot of unignorable problems such as noise robustness, interpretability, and treatments of discrete data.
Specifically, using simple correlations only, large input dimension causes difficulty of finding significant relationships because the reasonable threshold selection is not trivial.
In order to find the essential relationships between variables, the sparse modeling, such as LASSO (Least Absolute Shrinkage and Selection Operator)\cite{Tibishinari96}, is focused in these decades\cite{Igarashi16}.  
Graphical Lasso~\cite{friedman_sparse_2008} is a representative method which achives sparse covariance structure analysis.
Since an inverse covariance matrix corresponds to a partial correlation matrix with appropriate scaling,
we can discover robust and essential relationships of data by sparse covariance structure analysis.
However, because of the assumption of a Gaussian Graphical Model (GGM) for observed data,
Graphical Lasso can not treat count data.

In order to overcome this limitation of Graphical Lasso,
we propose a hierarchical Bayesian model for Poisson distributed observations with sparse covariance structure.
In our model, the latent variables, which indicate ``\emph{potential risks}'' of events, follow a Bayesian Graphical LASSO(BGL) \cite{wang2012bayesian}, and the occurrences of events follow the homogeneous Poisson processes. 
We apply the proposed model to spatial crime data analysis and investigate the effectiveness with numerical experiments.

\subsection{Graphical Lasso}
Graphical Lasso~\cite{friedman_sparse_2008} is a well-known and powerful model for sparse covariance structure analysis of GGM.
Let a zero-meaned data matrix $\bs{Y} = (\bs{y}_1, \ldots, \bs{y}_T) \in \mathbb{R}^{T \times A}$
follows a Gaussian distribution independently with given a precision matrix $\bs{\Omega} \in \mathbb{R}^{A \times A}$ denoting each component as $\omega_{ij}$ in $(i, j = 1, \dots, A)$.
\begin{align}
    p(\bs{Y} | \bs{\Omega}) = \prod^T_{t = 1} \mathcal{N}(\bs{y}_t | \bs{0}, \bs{\Omega}^{-1}).
\end{align}
The Graphical Lasso optimizes the following objective which consists of the term from maximizing likelihood estimator of $\bs{\Omega}$
under the $L_1$ penalty
\begin{align}
    \label{gl_objective}
    \bs{\Omega} = \argmax_{\bs{\Omega}' \in \bf{M}^+} \log(\mathrm{det} (\bs{\Omega}')) - \mathrm{tr}(\bs{Y}^\top \bs{Y} \bs{\Omega}') - \lambda \|\bs{\Omega}'\|_1,
\end{align} 
where $\bf{M}^+$ indicates the set of all $A \times A$ positive semi-definite matrices
and $\lambda$ is a regularization parameter.
Here, the $L_1$ norm is defined by
$\|\bs{\Omega}\|_1 = \sum_{ij} |\omega_{ij}|$.
Note that maximizing the objective of Graphical Lasso \eqref{gl_objective} is equivalent to
obtain the maximum a posteriori (MAP) estimator of the following model, which is 
the combination of the Laplace distributions of the off-diagonal components of $\Omega$ denoting as $\mathrm{DE}(\omega_{ij}\mid\lambda)$ of the form $p(\omega_{ij})=\lambda/2 \exp(-\lambda|\omega_{ij}|)$ and the exponential distribution of the diagonal components $\mathrm{Exp}(\omega_{ii}\mid \lambda/2)$ of the form $p(\omega_{ii})=\lambda/2 \exp(-\lambda\omega_{ii}/2)$:
\begin{align}
    \label{gl_map}
    p(\bs{Y} \mid \bs{\Omega}) &= \prod^T_{t = 1} \mathcal{N}(\bs{y}_t \mid \bs{0}, \bs{\Omega}^{-1}),\\
    p(\bs{\Omega} \mid \lambda ) &= C^{-1} \prod_{i<j} 
    \left \{
        \mathrm{DE}(\omega_{ij} |\,  \lambda )
    \right \} 
    \prod_{i=1}^{A} \mathrm{Exp}
    \left ( \omega_{ii} 
        \left |  \cfrac{\lambda}{2} 
        \right . 
    \right )
    1_{\bs{\Omega} \in \bs{M}^+},
\end{align}
where $C^{-1}$ is a normalizing term and $1_{\bs{\Omega} {\in \bs{M}^+}}$ is an indicator function defined by
\begin{align}
    1_{\bs{\Omega} \in \bs{M}^+} =
    \left \{
    \begin{array}{ll}
        1 &~~~ (\bs{\Omega} \in \bs{M}^+) \\
        0 &~~~ (\mbox{otherwise}).
    \end{array}
    \right .
\end{align}

\subsection{Bayesian Graphical Lasso}
Bayesian Graphical Lasso (BGL)~\cite{wang2012bayesian} realizes a fully Bayesian treatment of Graphical Lasso.
Using the fact that the double exponential distributions can be represented as a mixture of Gaussian and exponential distributions,
\eqref{gl_map} is modified to
\begin{align}
    \label{bgl_model}
   p(\bs{\Omega} | \bs{\tau}, \lambda ) &= C^{-1}_{\bs{\tau}} \prod_{i<j} \mathcal{N}(\omega_{ij} | 0, \tau_{ij})
   \prod_{i=1}^{A} \mathrm{Exp} \left( \omega_{ii} \left | \cfrac{\lambda}{2} \right . \right ) 1_{\bs{\Omega} {\in \bs{M}^+}},\\
   p(\bs{\tau} | \lambda) &\propto C_{\bs{\tau}} \prod_{i<j} \cfrac{\lambda ^ 2}{2} \exp \left (- \cfrac{\lambda ^ 2}{2} \tau_{ij} \right ),
\end{align}
where $\bs{\omega} = \{\omega_{ij}\}_{i \leq j}$ denote the vector of the upper off-diagonal and diagonal entries of
$\|\bs{\Omega}\|$ and $\bs{\tau} = \{\tau_{ij}\}_{i < j}$ is the latent scale parameters.
From the decomposition of the double exponential distributions,
we derive a data-augmented block Gibbs sampling algorithm for the BGL model\eqref{bgl_model}.

\subsection{The Poisson Process and Crime Data}
A Poisson distribution gives probability masses of the numbers of occurred events per unit time.
More precisely, if the time intervals of events' occurrences follow identical exponential distributions,
the numbers of events per unit time follow identical Poisson distributions.
This model is called a homogeneous Poisson process.
The Poisson process is a simple but reasonable model to represent occurrences for rare events.
Common examples of Poisson processes are customers calling a help ~\cite{weinberg2007bayesian}, radioactive decay in atoms~\cite{sitek2015limitations}, and crime occurrence~\cite{osgood2000poisson}, anomaly detection~\cite{ihler_adaptive_2006}, and so on.

If data follow a multivariate point process,
it is significant to understand the inter-variate relationships of data.
When the data consists of continuous values,
we can evaluate the structure of variables by calculating correlations or applying Graphical Lasso.
However, for count data, the correlation matrix will lead to biased evaluation for the structure.
Hence, we introduce a novel Bayesian framework of sparse covariance structure analysis for count data.

The remainder of this paper is organized as follows.
Section~\ref{met} gives the concrete formulation of the proposed model.
In Section~\ref{sim}, we show the effectiveness of our model through experiments with spatial crime data and Section~\ref{ana} shows the discussion of our model.
Finally, Section~\ref{con} is devoted to a summary of this study.\par

\section{The Proposed Method}
\label{met}
\begin{figure}[t]
    \centering
    \includegraphics[width = 8.0cm]{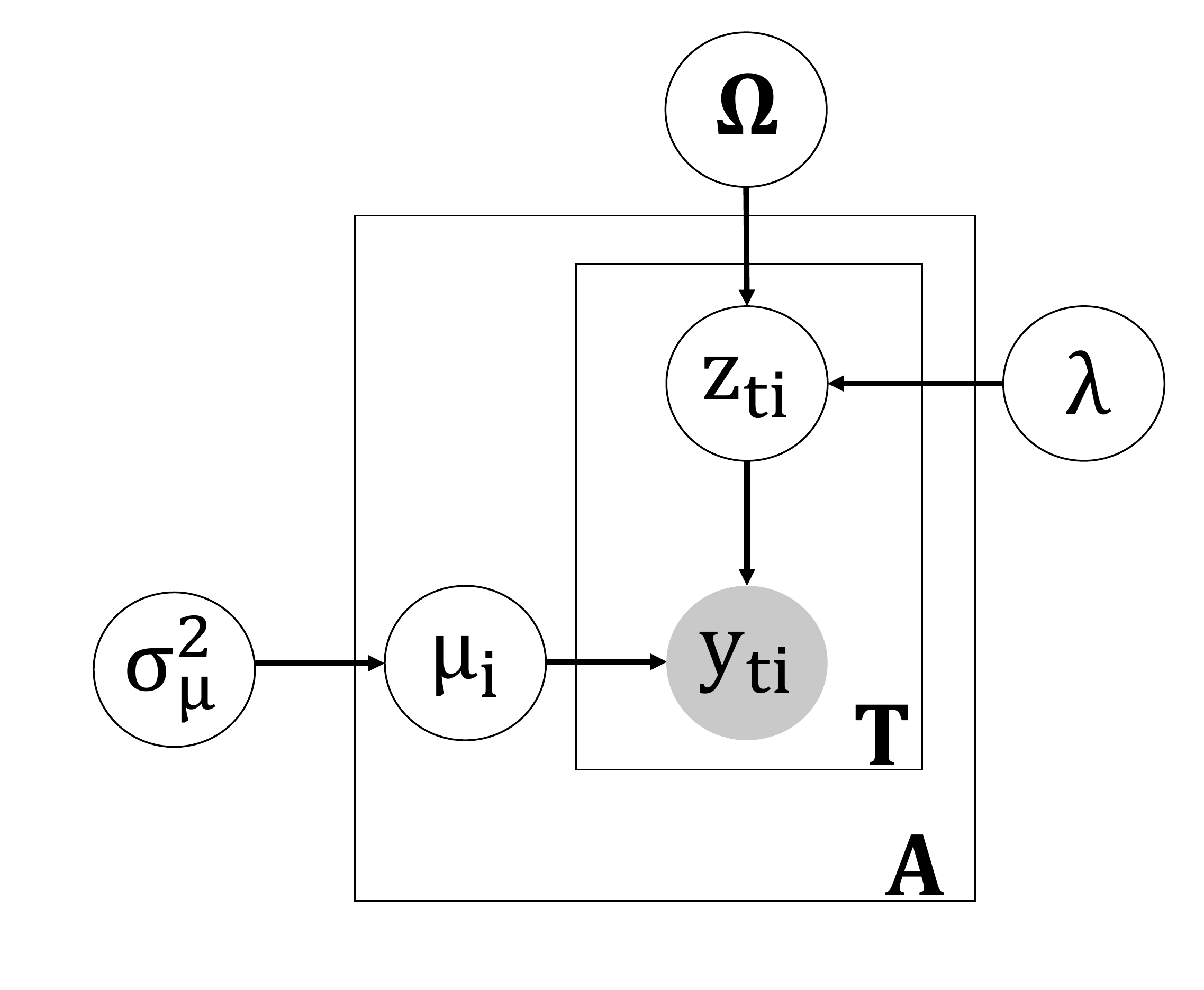}
    \caption{The graphical model of the proposed model}
    \label{fig:graphicalmodel}
\end{figure}

\subsection{Our Model}
Fig. \ref{fig:graphicalmodel} shows the graphical model of the proposed model.
At First, we define the set of non-negative integers $\mathbb{N}_0 = \mathbb{N} \cup \{0\}$.
When we obtain $A$-dimensional sequential data for $T$ timesteps,
we assume that the elements of count data matrix $\bs{Y} \in \mathbb{N}^{T \times A}_0$ follows
conditionally independent Poisson distributions
\begin{align}
    \label{likelihood}
   P(\bs{Y} |\bs{\mu}, \bf{Z}) &= \prod_{i=1}^{A} \prod_{t=1}^{T} \mathrm{Poisson}(y_{ti} | \exp(\eta(\mu_i, z_{ti}))),\\
   \eta(\mu_i, z_{ti}) &= \mu_i + z_{ti},
\end{align}
and we also give priors as
\begin{align}
    \label{prior_mu}
    p(\bs{\mu}) &= \mathcal{N}(\bs{\mu} | \bs{0}, \sigma^2_\mu \bs{I}),\\
    \label{prior_z}
    p(\bs{z}_t | \bs{\Omega}) &= \mathcal{N}(\bs{z}_t | \bs{0}, \bs{\Omega}^{-1}) ~~ \mbox{for} ~ t = 1, \ldots, T, \\
    p(\bs{\omega} | \lambda) &\propto \prod_{i<j} \mathrm{DE}(\omega_{ij} | \lambda ) \prod_{i=1}^{A} \mathrm {Exp} \left ( \omega_{ii} \left | \cfrac{\lambda}{2} \right . \right ) 1_{\bs{\Omega} \in \bs{M}^+},\\
    p(\lambda) &= \mathrm{Gamma}(\lambda | a_\lambda, b_\lambda).
\end{align}
Here, we assume the linear predictor $\eta(\mu_i, z_{ti}) = \mu_i + z_{ti}$ as the \textit{potential risk} of occurrence of events.
Thus, $\mu_i$ indicates averaged potential risk of $i$-th dimension and
$z_{ti}$ represents dispersities from $\mu_i$.
Since $\bs{z}_t$ follows the BGL prior, we can extract the sparse and essential co-occurrence structures of count data.
In addition, we discussion about the effects of choices of $p(\lambda)$ in Section~\ref{sim}.
Therefore, the joint posterior distribution can be expressed as
\begin{align}
    \label{posterior}
   p(\bs{Z}, \bs{\mu}, \bs{\Omega}, \lambda | \bs{Y}) \propto P(\bs{Y}|\bs{\mu}, \bs{Z}) 
   p(\bs{\mu}) \left [ \prod^T_{t=1} p(\bs{z}_t | \bs{\Omega}) \right ] p(\bs{\omega} | \lambda) p(\lambda).
\end{align}

\subsection{Sampling Scheme}

We evaluate the posterior \eqref{posterior} with a Markov chain Monte Carlo method which consists of two different sampling schemes.
The first scheme is block Gibbs sampling for the parameters in the BGL, {\it{i.e.}} $\bs{\Omega}$ and $\lambda$.
The second one is about the parameters of the potential risks, that is, $\bs{\mu}$ and $\bs{Z}$.
We adopt the Metropolis-Hastings scheme for correlated count data~\cite{chib_marginal_2001}.\\
For example, we describe the Metropolis-Hastings sampling scheme from the fully conditional distribution of $\bs{\mu}$ here.
When $\bs{\mu}$ tries to transition to new state $\bs{\mu}'$,
we define the acceptance probability $r$ for simplicity
\begin{align}
    r = \min \left \{ 1, \cfrac{p(\bs{\mu}' | \bs{Y}, \bs{Z}, \bs{\Omega}, \lambda)p(\bs{\mu} | \bs{\mu}')}{p(\bs{\mu} | \bs{Y}, \bs{Z}, \bs{\Omega}, \lambda)p(\bs{\mu}' | \bs{\mu})} \right \}
\end{align}
The sampling procedure is as follows:
\begin{description}
\item[Step 1]\mbox{}\\
Find optimal $\bs{\mu}$ about the fully conditional distribution
\begin{align}
    \label{fully_conditional_mu}
    \hat{\bs{\mu}} = \argmax_{\bs{\mu}} \log p(\bs{\mu} | \bs{Y}, \bs{Z}, \bs{\Omega}, \lambda) 
\end{align}
by use of the Newton-Raphson method.
\item[Step 2]\mbox{}\\
Sample a candidate state $\bs{\mu}'$ from proposal distribution $p(\bs{\mu}' | \hat{\bs{\mu}})$ defined as the multivariate t-distribution:
\begin{align}
    p(\bs{\mu}'|\hat{\bs{\mu}}) = \textrm{Multi-t}(\bs{\mu}' | \hat{\bs{\mu}}, \bs{H}_{\hat{\mu}}^{-1}, \nu),
\end{align}
where $\bs{H}_{\mu} \in \bs{R} ^ {A \times A}$ is the Hessian matrix of \eqref{fully_conditional_mu} and
$\nu (> 0)$ is a user-defined hyperparameter which indicates the degree of freedom.
\item[Step 3]\mbox{}\\
Accept the candidate state $\bs{\mu}'$ with probability $r$.\\
Update the state $\bs{\mu} \gets \bs{\mu}'$ if accepted, and keep the state to be if rejected.
\end{description}

Sampling from $p(\bs{z}_t | \bs{Y}, \bs{Z}_{\backslash t}, \bs{\mu}, \bs{\Omega}, \lambda)$ is also executed by the similar way to $\bs{\mu}$.
In summary, we repeat sampling from the fully conditional distributions of $\bs{\Omega}, \lambda, \bs{\mu}$ and $\bs{Z}$
respectively with Metropolis-Hastings within Gibbs sampler.

\section{Synthetic Data Analyses}
\label{sim}
\subsection{Synthetic Data}
To assess the performance of our proposed model, we generate four synthetic datasets,
whose size are $(A, T) = (10, 30), (50, 60), (100, 60), (200, 60)$. 
We fix $\mu_i = 0.2 ~~ (i = 1, \ldots, A)$ and generated $\bs \Omega$ as $\bs{M}^+$ with $\omega_{ii} = C_1$, 
$\omega_{i,i-A/2} = \omega_{i-A/2,i} = C_2$ and zero otherwise for entire simulations, where $C_1$ and $C_2$ represent constant values.
Given these true parameters, we sample $\bs{Z}$ and the observed data matrix $\bs{Y}$ from \eqref{likelihood} and \eqref{prior_z} respectively.

\subsection{Analyzing Effects of Hyperparameter Selection}
In our Metropolis-Hastings sampling scheme,
we empirically find that an appropriate selection for the degree of freedom $\nu$ in proposal t-distribution gives
a significant effect for the sampling efficiency.
In our simulations, slightly small $\nu$, such as $\nu = 5$, is better than bigger $\nu$.
Because a t-distribution with $\nu = 1$ is equivalent to Cauchy distribution and
$\nu \to \infty$ is corresponding to a Gaussian distribution,
our choice $\nu = 5$ means intermediate form of them.

Next, we determine the parameters in priors.
For $a_\lambda$, which is the hyperparameter of the regularization parameter $\lambda$,
we adopt $a_\lambda = A$ for $A = 10, 50, 100$ and $a_\lambda = 0.01$.
Moreover, a value $\sigma^2_{\mu} = 0.05$ seems to be reasonable
for the prior $p(\bs{\mu})$.
Since $\bs{\mu}$ and $\bs{Z}$ determine the parameters of the Poisson distribution within the exponential function in the proposed model, the absolute value of $\mu$ should be small.

\subsection{Simulation Results}
Here we show the estimation results of $\bs{\mu}$, $\bs{Z}$, $\bs{\Omega}$
in the case that the size of data is $(A, T) = (50,60)$.
We adopt the hyperparameters $(a_\lambda, \sigma^2_{\mu}, \nu) = (A, 0.05, 3)$ and
updated all elements at once both $\bs{\mu}$ and $\bs{z}_t$ in the Metropolis-Hastings algorithms.

\begin{figure}[t]
    \centering
    \includegraphics[width = 9.0cm]{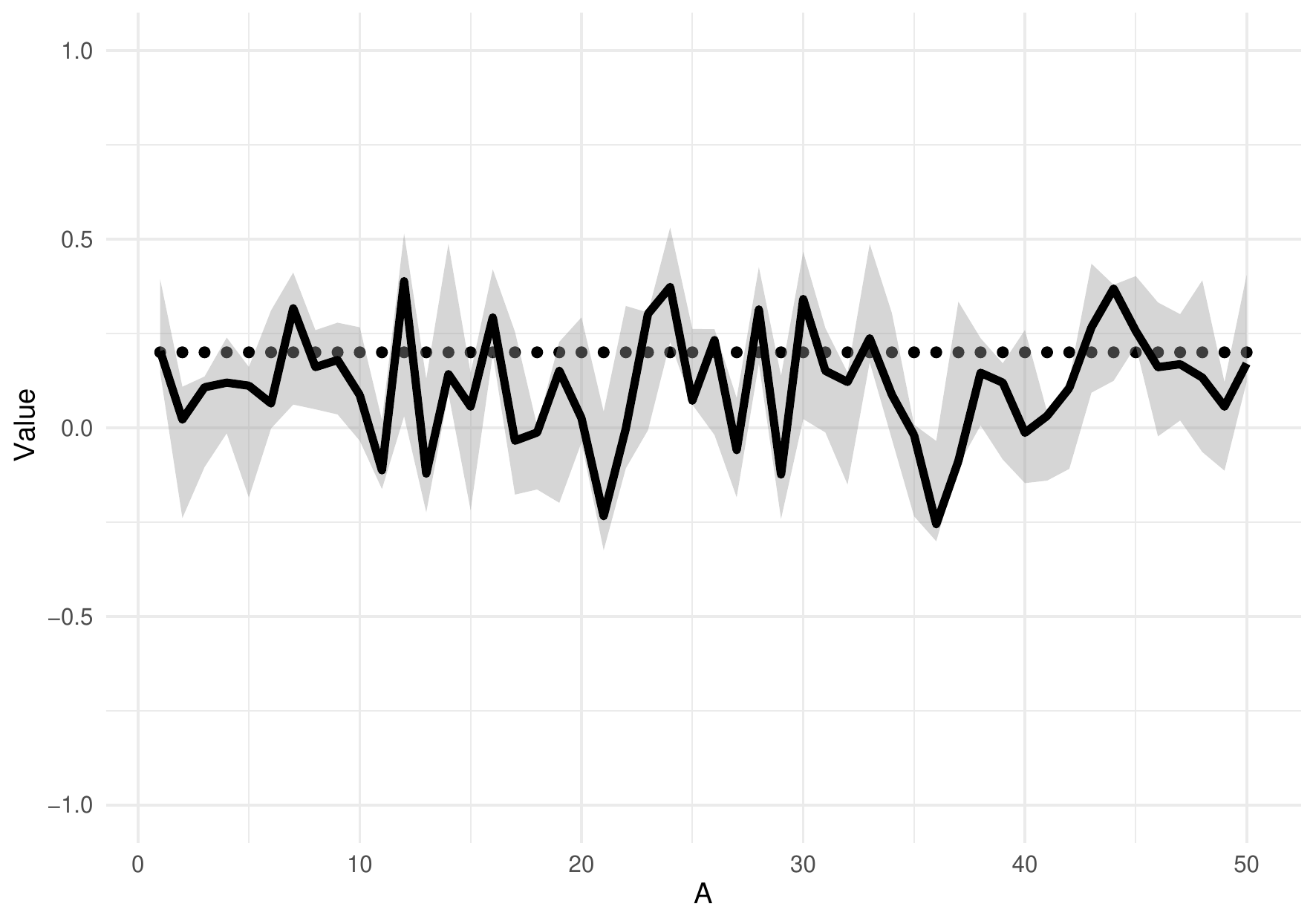}
    \caption{The estimation result of $\bs{\mu}$ with the MAP estimator with 95\% credible intervals.}
    \label{fig:mu_HPD}
\end{figure}
Fig. \ref{fig:mu_HPD} shows the result of $\bs{\mu}$.
We show the solid line estimates of $\mu_i$ with the MAP estimator
with 95\% credible intervals.
The dotted line represents the true values.
It is assumed that small $\mu_i$ is difficult to estimate
because the linear predictor $\eta(\mu_i, z_{ti})$ is exponentially transformed to be the parameter of Poisson distribution.
However, we find the 95\% the true values are held within credible intervals.
\begin{figure}[t]
    \centering
    \begin{minipage}[cl]{0.46\columnwidth}
        \centering
        \includegraphics[width = \columnwidth]{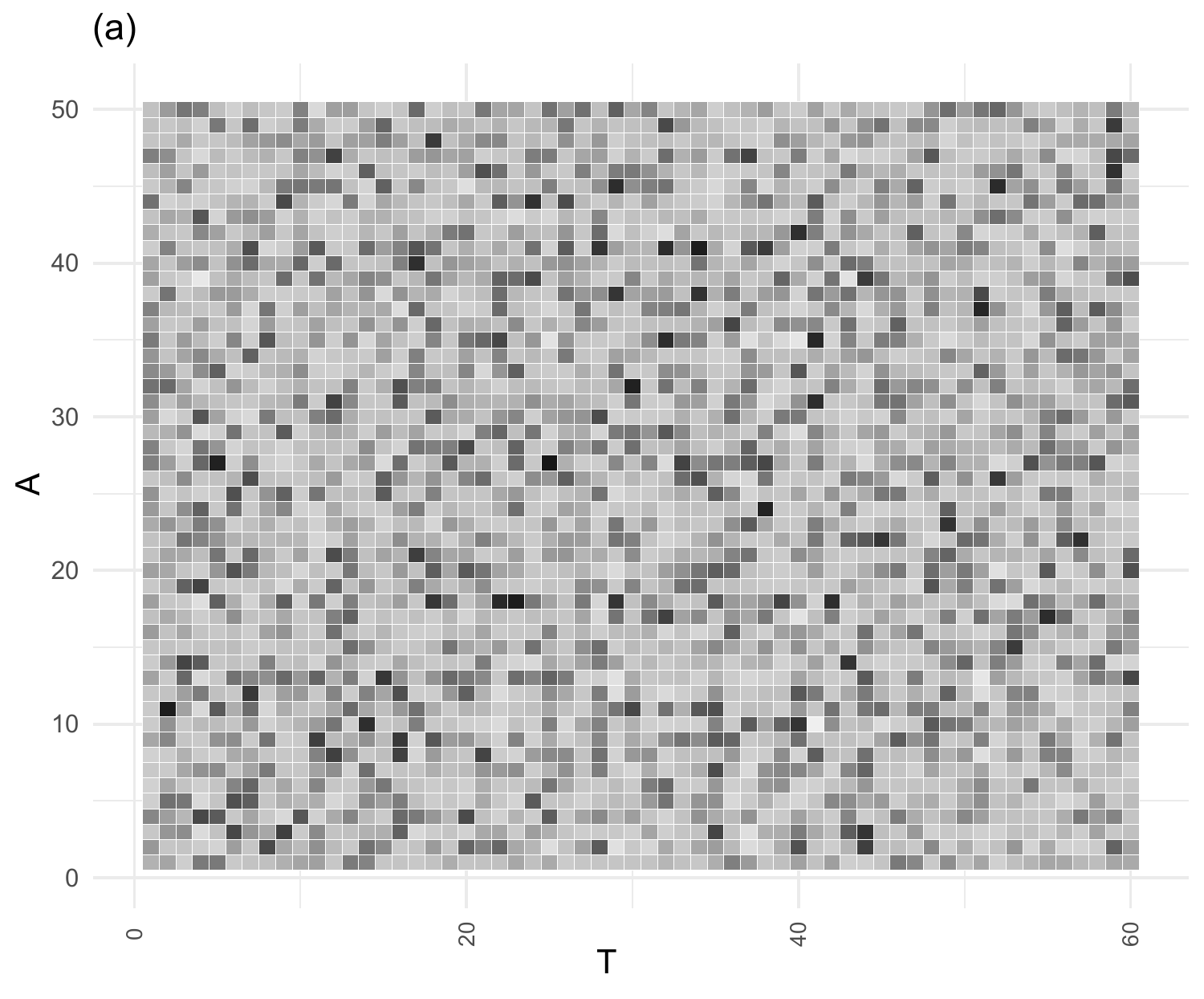}
    \end{minipage}
    \begin{minipage}[cl]{0.46\columnwidth}
        \centering
        \includegraphics[width = \columnwidth]{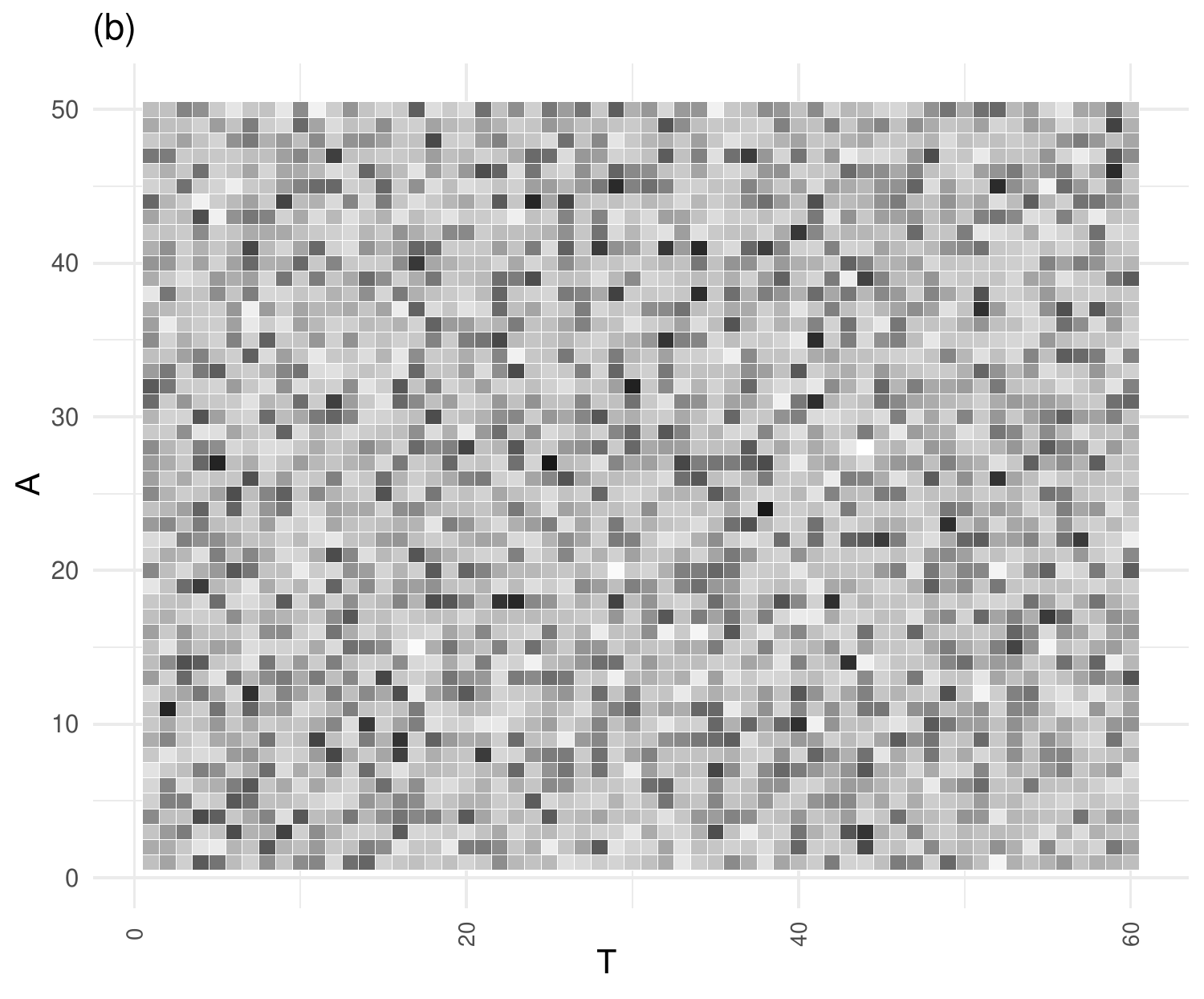}
    \end{minipage}
    \begin{minipage}[cl]{0.06\columnwidth}
        \centering
        \includegraphics[width = \columnwidth]{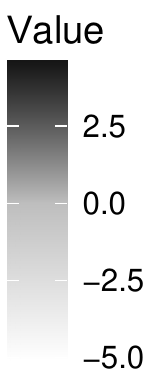}
    \end{minipage}\\
    \hspace{-20mm}
    \begin{minipage}[cl]{0.46\columnwidth}
        \centering
        \includegraphics[width = \columnwidth]{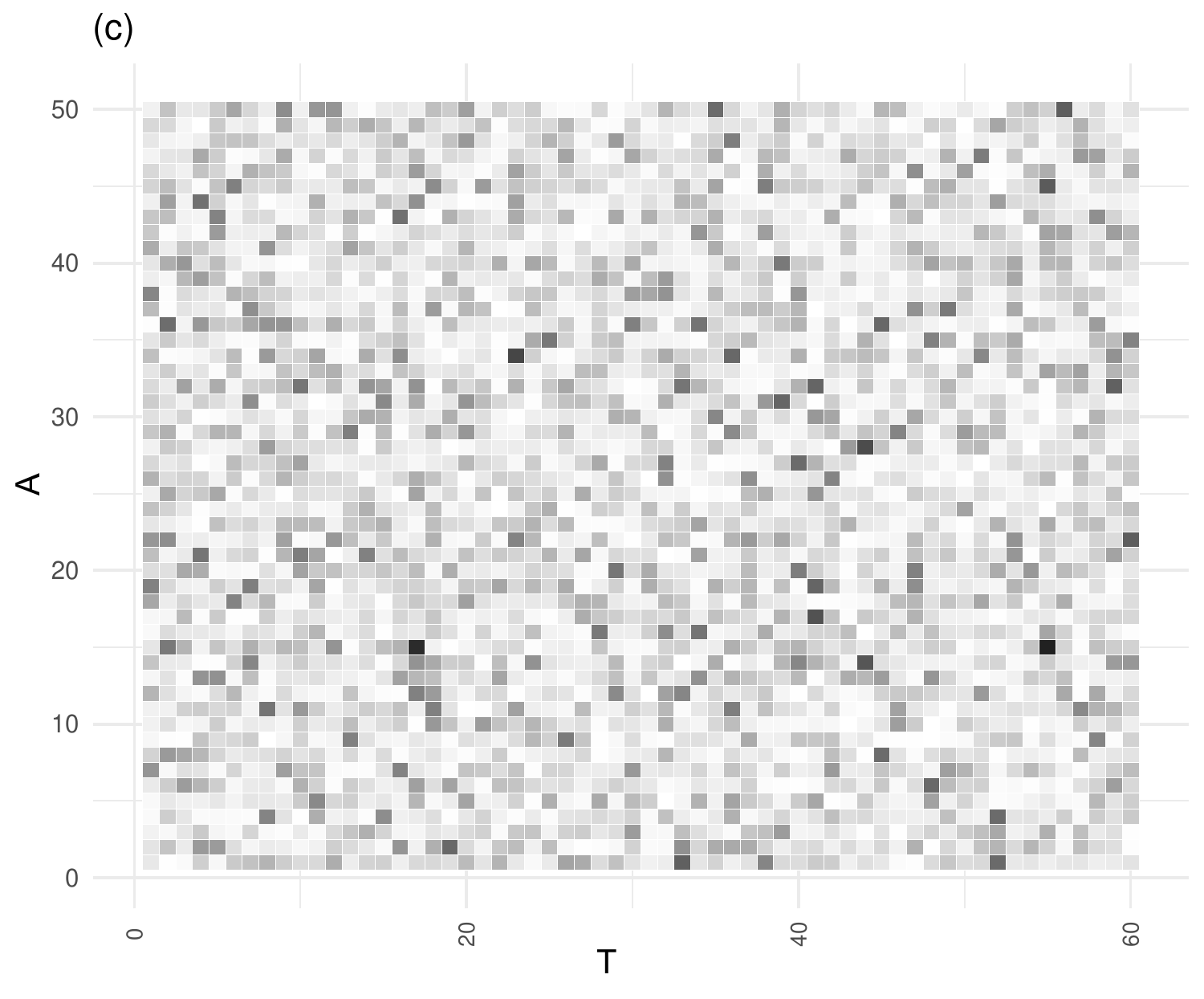}
    \end{minipage}
    \hspace{-20mm}
    \begin{minipage}[cl]{0.52\columnwidth}
        \centering
        \includegraphics[width = 0.1\columnwidth]{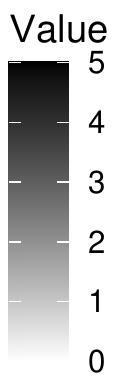}
    \end{minipage} 
    \caption{(a) shows the MAP estimation results of $\bs{Z}$, (b) is true value of $\bs{Z}$, and (c) shows the difference between (a) and (b) in absolute value.}
    \label{fig:Z}
\end{figure}
Fig. \ref{fig:Z} shows the MAP estimation results of $\bs{Z}$ and their true values.
The subfigure (c) shows the differences between MAP estimated and true $\bs{Z}$ in absolute value.
\begin{figure}[t]
    \centering
    \includegraphics[width = 10.0cm]{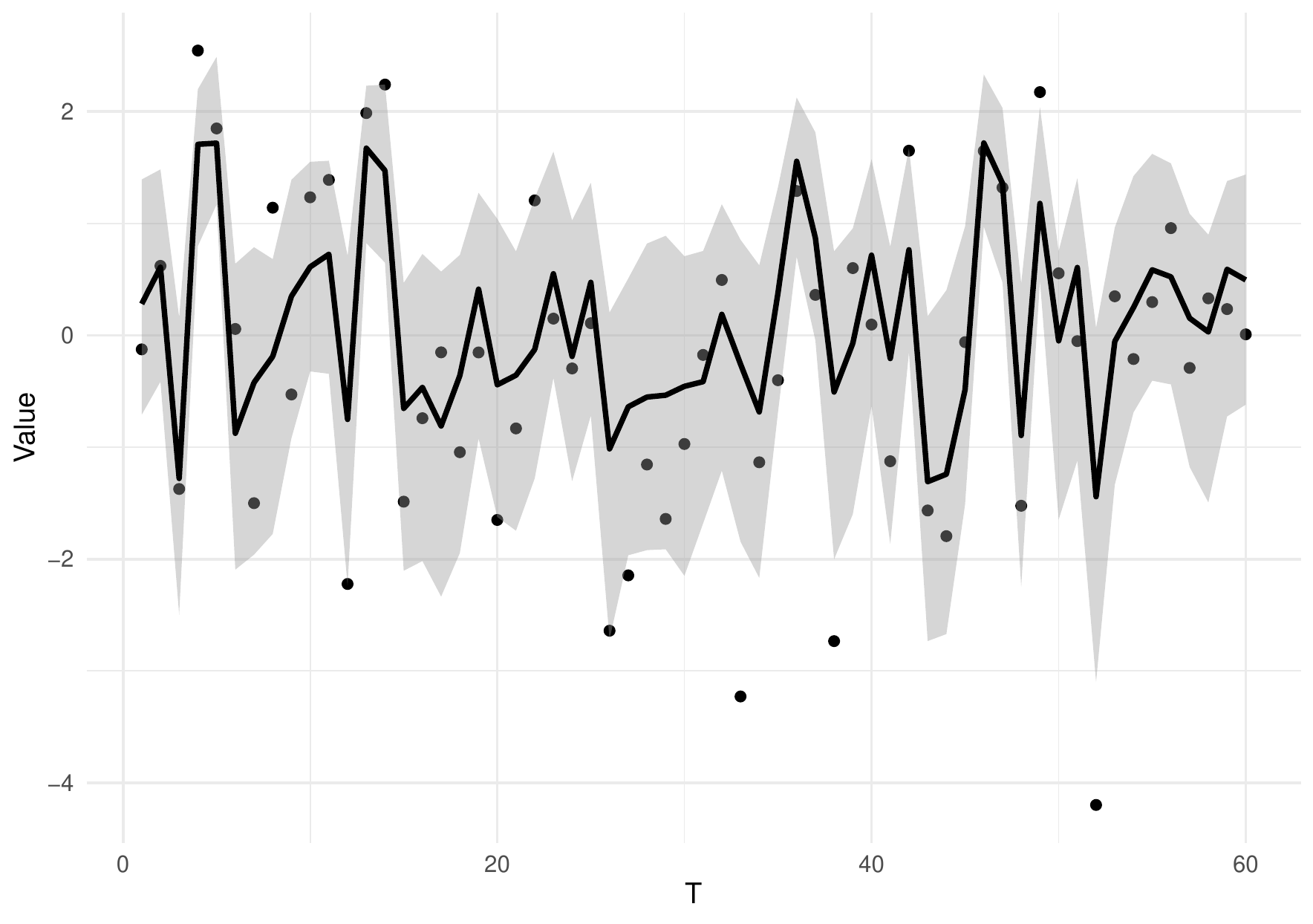}
    \caption{The estimation result of $(z_{t1})^T_{t = 1}$. The markers show the true values of and the solid line shows MAP estimated values with the 95\% credible intervals.}
    \label{fig:Z_HPD}
\end{figure}
Fig. \ref{fig:Z_HPD} shows the estimation result of the value of $(z_{t1})^T_{t = 1}$.
Note that $\mu_1$ is estimated to be $0.2$ as shown in Fig. \ref{fig:mu_HPD}.
The solid line is the MAP estimation results of $(z_{t1})^T_{t = 1}$ with the 95\% credible intervals,
and the markers are their true values.
For the positive $z_{t1}$, the true values are within the credible intervals of the estimation result.
On the other hand, the true values seem to be out of the 95\% credible intervals for the negative $z_{t1}$.
However, considering that the parameter of Poisson distribution becomes a small value,
the estimation results are reasonable.

\subsection{Evaluating Partial Correlations}
\begin{figure}[t]
     \begin{tabular}{cc}
      \begin{minipage}[c]{0.80\hsize}
      \centering
    \includegraphics[width = 9.0cm]{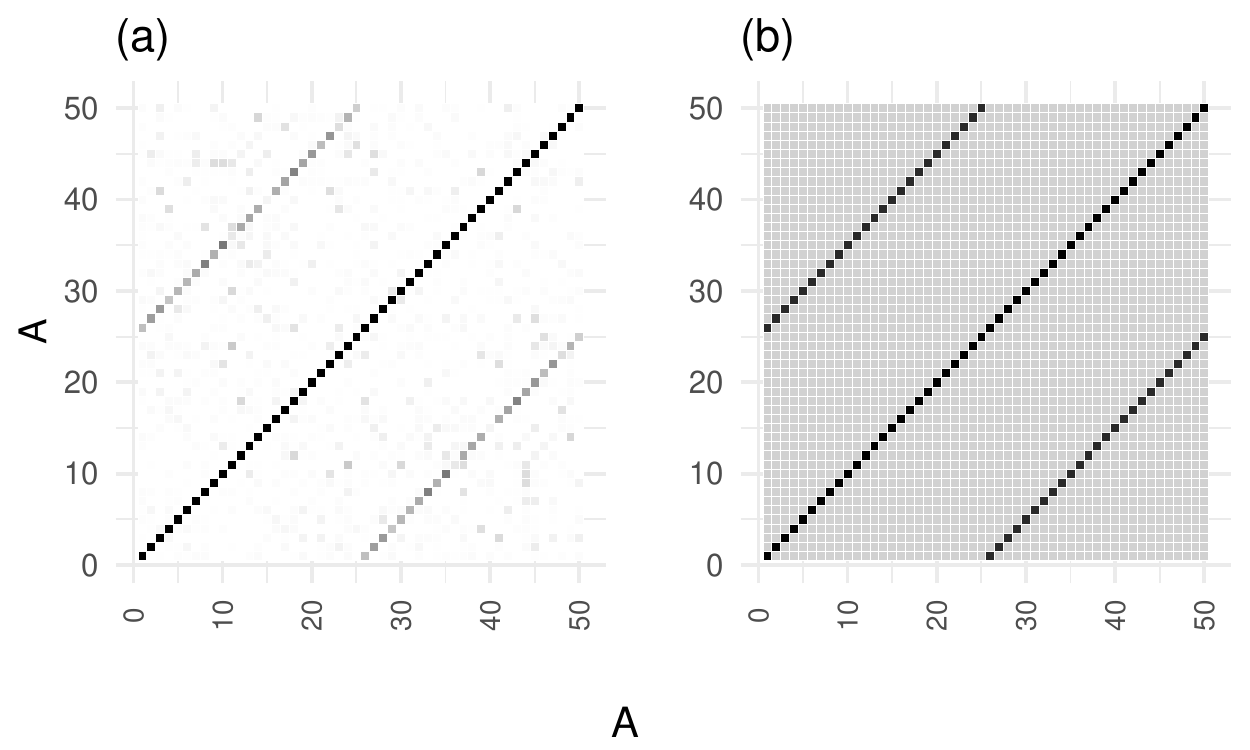}
     \end{minipage} &
     \begin{minipage}[c]{0.15\hsize}
     \centering
    \includegraphics[keepaspectratio, scale=0.30, height=20mm]{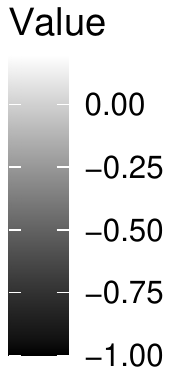}
     \end{minipage}
    \end{tabular}
    \caption{(a) shows the estimated partial correlation coefficients of $\bs\Omega$ and  (b) shows the true partial correlation coefficients of $\bs\Omega$. }
    \label{fig:pcor}
\end{figure}
We can calculate partial correlation matrix $\bs{P}$ as
\begin{align}
    \label{eq:pcor}
    p_{ij} = - \cfrac{\omega_{ij}}{\sqrt{\omega_{ii}}\sqrt{\omega_{jj}}}
\end{align}
from the estimated precision matrix $\bs{\Omega}$.
In Fig. \ref{fig:pcor}, we show the MAP estimated $\bs{P}$ for $(A, T) = (50, 60)$.
%
%
We compare the estimated partial correlation coefficients and true one, which consists of small number of non-zero components.
We confirm the partial correlation coefficients that correspond to the non-zero components are relatively larger than the other components that correspond to the zero components in the true partial correlation coefficients.
%
We also find that non-diagonal and non-zero estimates have shrinkage from their true values.
This is affected by the Lasso-like priors of BGL.

\section{Analysis Of Crime Spots Data}
\label{ana}
\subsection{Spatial Crime Data}
As an application example of the proposed model,
we employed the proposed model to spatial crime data
which is obtained from~\cite{homepage_NIJ}.
The spatial crime data published by the National Institute of Justice (NIJ)
contains criminal occurrences and their locations in the Portland City, Oregon, USA.

In the crime data, the latent variable $\mu_i$ represents the average potential risk of criminal occurrences of the $i$-th area.
Also, $\bs{Z}$ represents dispersities of the potential risks at each time point.
Therefore, the interaction structure is captured by $\bs{Z}$.

For simplicity, we used the crimes that occurred in 2016 and extracted 60 areas.
Furthermore, we aggregate the number of crimes per week into one time point.
Hence, the size of data matrix becomes $\bs{Y}$ is $(A, T) = (60, 52)$.

\subsection{The Partial Correlation Of $\bs \Omega$ On Crime Data}
The estimated $\bs \Omega$ is the inverse of the sparse covariance matrix that contributes to the influence $\bs Z$ between variables and temporal variation.
Therefore, the partial correlation calculated from $\bs \Omega$ represents a sparse correlation of crime occurrence risk between areas.

A strong positive correlation between two areas means that the risk of the crime in one area intends to increase when the other area risk increases.

\subsection{Visualization Of Partial Correlations}
\begin{figure}[t]
    \centering
    \includegraphics[width = 10.0cm]{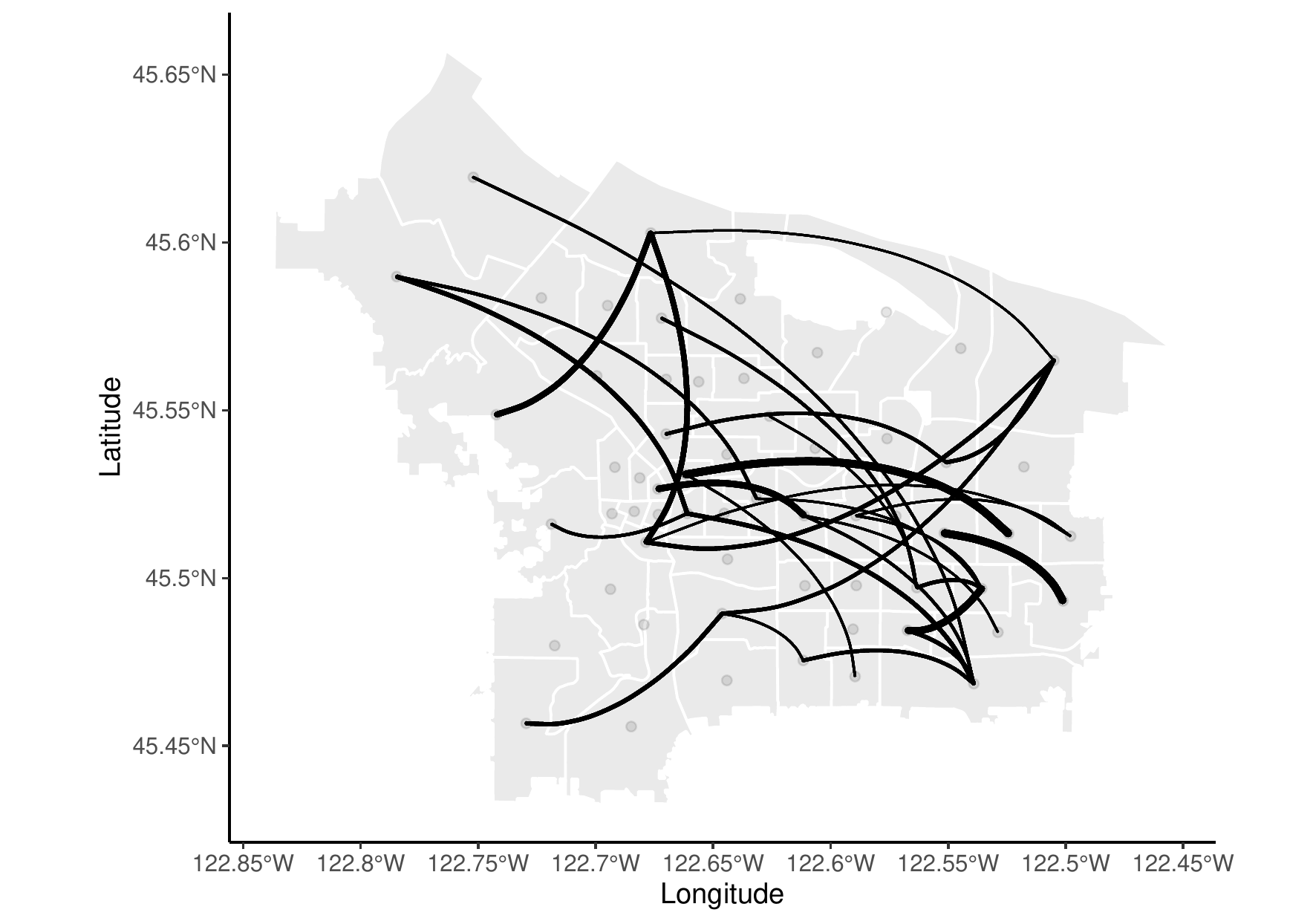}
    \caption{The visualization of sparse partial correlation coefficients between areas on Portland city's map. The thickness of the curved black lines shows the magnitudes of the coefficients.}
    \label{fig:map}
\end{figure}
The partial correlation coefficients matrix is calculated from \eqref{eq:pcor} with the estimated result of $\bs \Omega$ .
Fig. \ref{fig:map} shows a visualized sparse partial correlation coefficients on the Portland city map. The coefficients shown on the map are the top 2\% of the whole in absolute value.
The correlation of crime risks could lead to a visual understanding by expressing a strong correlation on the map.

\section{Conclusion}
\label{con}
The proposed model can estimate reasonable values of $\bs \Omega$ and latent variables $\bs \mu$ and $\bs Z$.
On the other hand, when the parameter is less than 1 in the Poisson process, 
the number of events tends to be $0$, so that it is difficult to estimate the negative true value of the latent variable of the simulation data in the proposed model.
One of the possible solutions tackle this problem is to find an alternative transformation function which
maps $\mathbb{R}$ to $[0, \infty)$ for the linear predictor of Poisson distribution. 
For example, $x \mapsto \log ( 1 + \exp(x))$ is one of the candidates.

It is possible to find useful features by analyzing the estimation results of latent variables by the proposed model.
We have been able to obtain sparse correlations between crime risk areas, by applying the proposed model to crime data.
As a future issue, because there are many samples related to time in the Poisson process in general,
we would like to make our model possible to catch time-series dependence.

%
%

\bibliographystyle{splncs04}
\bibliography{main}

\end{document}